\documentclass[
 reprint,
 amsmath,
 amssymb,
 aps
]{revtex4-2}

\usepackage{graphicx}
\usepackage{dcolumn}
\usepackage{bm}

\DeclareMathAlphabet{\pazocal}{OMS}{zplm}{m}{n}

\def\imo{i}

\def\K{{\cal K}}

\begin{document}

\title{Bardeen spacetime as a quantum corrected Schwarzschild black hole:\\ Quasinormal modes and Hawking radiation}

\author{R. A. Konoplya}
\email{roman.konoplya@gmail.com}
\author{D. Ovchinnikov}
\email{dmitriy.ovchinnikov@physics.slu.cz}
\affiliation{Research Centre for Theoretical Physics and Astrophysics, Institute of Physics, Silesian University in Opava, Bezručovo náměstí 13, CZ-74601 Opava, Czech Republic}

\author{B. Ahmedov}
\email{ahmedov@astrin.uz }
\affiliation{Ulugh Beg Astronomical Institute, Astronomy St. 33, Tashkent 100052, Uzbekistan}
\affiliation{Institute of Fundamental and Applied Research, National Research University TIIAME, Kori Niyoziy 39, Tashkent 100000, Uzbekistan}
\affiliation{National University of Uzbekistan, Tashkent 100174, Uzbekistan}

\date{\today}

\begin{abstract}
The Bardeen black hole holds historical significance as the first model of a regular black hole. Recently, there have been proposed interpretations of the Bardeen spacetime as quantum corrections to the Schwarzschild solution. Our study focuses on investigating the quasinormal modes and Hawking radiation of the Bardeen black hole. We have observed that previous studies on the quasinormal modes for the Bardeen black hole suffer from inaccuracies that cannot be neglected. Therefore, we propose accurate calculations of the quasinormal modes for scalar, electromagnetic, and neutrino fields in the Bardeen spacetime. Additionally, we have computed the grey-body factors and analyzed the emission rates of Hawking radiation. Even when the quantum correction is small and the fundamental mode only slightly differs from its Schwarzschild value, the first several overtones deviate at an increasingly stronger rate. This deviation leads to the appearance of overtones with very small real oscillation frequencies. This outburst of overtones is closely linked to the fact that the quantum-corrected black hole differs from its classical limit primarily near the event horizon. Moreover, the intensity of the Hawking radiation is significantly suppressed (up to three orders of magnitude) by the quantum correction.
\end{abstract}

\maketitle


\section{\label{sec:Introduction}Introduction}

The problem of a central singularity in the Schwarzschild solution represents a limitation of Einstein's theory of gravity. Consequently, the development of regular black hole models within various approaches and theories of gravity has become an important endeavor. The Bardeen black hole stands as the first historically significant ad hoc model describing a regular black hole \cite{Bardeen}. Subsequently, an interpretation of the Bardeen metric was proposed in the context of a specific nonlinear electrodynamics in curved spacetime, positing it as a giant gravitating magnetic monopole \cite{Ayon-Beato:2000mjt}. However, if the weak field limit is reduced to the usual Maxwell electrodynamics, the electric charge must vanish, as per the Bronnikov theorem \cite{Bronnikov:2000vy}. Despite such an exotic interpretation of the Bardeen spacetime, numerous studies have been dedicated to exploring various effects around magnetic monopoles, particularly their characteristic oscillation frequencies known as \textit{quasinormal modes} \cite{Fernando:2012yw, Breton:2016mqh, Flachi:2012nv, Toshmatov:2015wga, Toshmatov:2019gxg, MahdavianYekta:2019pol, Liu:2020ddo, Mondal:2020pop, Rincon:2020cos, Lopez:2022uie, Ulhoa:2013fca, Macedo:2016yyo, Wahlang:2017zvk, Saleh:2018hba, Dey:2018cws, Jusufi:2020agr, Rayimbaev:2022mrk, Wu:2022tak, Sun:2023woa, Vishvakarma:2023csw, Liu:2023kxd}.

Quasinormal modes not only serve as fundamental observables that characterize black holes \cite{LIGOScientific:2016aoc}, but also play a crucial role in gauge/gravity duality for describing strongly coupled quantum systems \cite{Son:2007vk}, as well as in determining the (in)stability of black holes \cite{Konoplya:2011qq}.

However, a significant portion of the aforementioned quasinormal mode calculations for the Bardeen black hole suffer from considerable numerical inaccuracies due to the utilization of various approximations. Consequently, it becomes necessary to reexamine some of these results using accurate methods based on converging procedures, such as the Leaver methods \cite{Leaver:1985ax}.

Nevertheless, the focus of our investigation here is on the Bardeen spacetime not within the context of an exotic nonlinear electrodynamics, but rather as a quantum-corrected solution to the Schwarzschild spacetime. In this framework, the parameter that formerly represented the magnetic charge now controls the magnitude of the quantum correction for a neutral black hole. Remarkably, stringy corrections to Schwarzschild black hole spacetimes arising from string T-duality were proposed in \cite{Nicolini:2019irw}. As an initial step, the static Newtonian potential was derived by exploiting the relationship between T-duality and path integral duality. Subsequently, it was demonstrated that the intrinsic non-perturbative nature of stringy corrections introduces an ultraviolet cutoff known as the zero-point length, resulting in a regular static potential. This finding was employed to derive a consistent black hole metric for a spherically symmetric, electrically neutral regular black hole, which is equivalent to the Bardeen spacetime after redefinition of constants. It is noteworthy that a similar duality is utilized to probe the quantum fluctuations of the background spacetime \cite{Padmanabhan:1998yya}. Ultimately, the Bardeen spacetime can also be deduced as an effective metric that reproduces the thermodynamics of a black hole within the Generalized Uncertainty Principle \cite{Maluf:2018ksj}, which represents another, more speculative approach to uncovering quantum corrections.

In this paper, we will investigate the quasinormal modes of scalar, electromagnetic, and neutrino fields in the background of the Bardeen spacetime, considering its aforementioned interpretation as a neutral quantum-corrected black hole. Our aim is to demonstrate the inaccuracies present in some of the previous calculations within certain parameter ranges. Here, we present precise calculations not only for the fundamental mode but also for the initial overtones with $\ell \leq n$, where $\ell$ represents the multipole number and $n$ denotes the overtone number. While it is commonly believed that the fundamental mode suffices as a fingerprint of the black hole, this is not entirely accurate. The fundamental mode is insensitive to the geometry of the event horizon and is primarily determined by the geometry near the peak of the effective potential. Consequently, if a black hole were to be replaced by a wormhole with the same geometry near the potential's peak, the fundamental mode would exhibit minimal change \cite{Damour:2007ap}. To ascertain the near-horizon behavior, it is necessary to consider multiple overtones \cite{Konoplya:2022pbc}, which also characterize the early phase of the ringdown \cite{Giesler:2019uxc}. Therefore, studying the initial few overtones will enable us to probe the geometry of the black hole in the vicinity of the event horizon.

We will demonstrate that the overtones deviate from their Schwarzschild limits at a significantly higher rate than the fundamental mode, and this deviation stems from the deformation of the metric near the event horizon \cite{Konoplya:2022pbc}. Similar outbursts have recently been observed in the context of black holes in higher-curvature corrected theories and asymptotically safe gravity \cite{Konoplya:2023ppx, Konoplya:2023aph, Konoplya:2022iyn, Konoplya:2022hll}.

Quantum-corrected black holes are expected to undergo intense Hawking evaporation, making it crucial to investigate Hawking radiation in the vicinity of Bardeen black holes. To the best of our knowledge, a comprehensive study of Hawking evaporation in this context has not yet been proposed.

In this paper, we will explore Hawking radiation of massless Standard Model fields around Bardeen black holes and demonstrate that the quantum correction significantly reduces the energy emission rate of the test fields (up to three orders of magnitude). To determine the emission rates, we will solve the problem of field scattering around the black hole, specifically by calculating the grey-body factors that account for the reflection from the effective potential and diminish the total emission rate.

The paper is organized as follows. In Section II, we provide a brief review of the Bardeen solution and the corresponding wave-like equations with effective potentials. In Section III, we describe the numerical methods employed to calculate quasinormal modes and grey-body factors, including the WKB approach, accurate Frobenius method, and time-domain integration. Section IV is dedicated to discussing the obtained quasinormal modes and the outburst of overtones. In Section V, we present the results of the calculations for grey-body factors and energy emission rates in Hawking radiation. Finally, in the Conclusions section, we summarize the obtained results and mention some open problems.

\section{\label{sec:Bardeen}Bardeen spacetime and the wavelike equations}
The Bardeen spacetime is described by the following line element 
\begin{equation}
		d s^2 = - f(r) d t^2 + f^{-1}(r) d r^2 + r^2 (d \theta^2 + \sin^2 \theta d \varphi^2),
\end{equation}
where
\begin{equation}
f(r) =  1 - \frac{2M r^2}{ ( r^2 + l_{0}^2 ) ^{3/2}}.
\end{equation}
For $l_{0}\neq 0$, the space-time in eq.(5) has horizons only if, $$ |l_{0}| \leq \frac{ 4 M}{ 3 \sqrt{3}},$$ as was shown in \cite{Borde:1994ai}. The parameter $l_{0}$ is related to the Regge slope as follows:
\begin{equation}
\sqrt{\alpha'} = \frac{l_{0}}{2 \pi} \approx 0.117 l_{P}, 
\end{equation}
where $l_{P}$ is the Plank mass, so that, as in \cite{Padmanabhan:1998yya}, we have:  $l_{0} \sim l_{P}$. 
Stringy effects produce a de Sitter core at the origin, because 
\begin{equation}
f(r) \rightarrow 1 -  \frac{\Lambda_{eff} r^{2}}{3},  \qquad  r \ll l_{0},
\end{equation}
with an effective cosmological constant 
$\Lambda_{eff} = 6 M/l_{0}^{3}$. This repulsive effect at small distances stabilizes the matter configuration against collapse.

The general relativistic equations for the scalar ($\Phi$), electromagnetic ($A_\mu$), and Dirac ($\Upsilon$) fields can be written as follows:
\begin{subequations}\label{coveqs}
\begin{eqnarray}\label{KGg}
\frac{1}{\sqrt{-g}}\partial_\mu \left(\sqrt{-g}g^{\mu \nu}\partial_\nu\Phi\right)&=&0,
\\\label{EmagEq}
\frac{1}{\sqrt{-g}}\partial_{\mu} \left(F_{\rho\sigma}g^{\rho \nu}g^{\sigma \mu}\sqrt{-g}\right)&=&0\,,
\\\label{covdirac}
\gamma^{\alpha} \left( \frac{\partial}{\partial x^{\alpha}} - \Gamma_{\alpha} \right) \Upsilon&=&0,
\end{eqnarray}
\end{subequations}
where $F_{\mu\nu}=\partial_\mu A_\nu-\partial_\nu A_\mu$ is the electromagnetic tensor, $\gamma^{\alpha}$ are gamma matrices and $\Gamma_{\alpha}$ are spin connections in the tetrad formalism.
After separation of the variables the above equations (\ref{coveqs}) take the Schrödinger wave-like form:
\begin{equation}\label{wave-equation}
\dfrac{d^2 \Psi}{dr_*^2}+(\omega^2-V(r))\Psi=0,
\end{equation}
where the ``tortoise coordinate'' $r_*$ is defined as follows:
\begin{equation}
dr_*\equiv\frac{dr}{f(r)}.
\end{equation}

The effective potentials for the scalar ($s=0$) and electromagnetic ($s=1$) fields have the form
\begin{equation}\label{potentialScalar}
V(r)=f(r) \frac{\ell(\ell+1)}{r^2}+\left(1-s\right)\cdot\frac{f(r)}{r}\frac{d f(r)}{dr},
\end{equation}
where $\ell=s, s+1, s+2, \ldots$ are the multipole numbers.
For the Dirac field ($s=1/2$) one has two iso-spectral potentials
\begin{equation}
V_{\pm}(r) = W^2\pm\frac{dW}{dr_*}, \quad W\equiv \left(\ell+\frac{1}{2}\right)\frac{\sqrt{f(r)}}{r}.
\end{equation}
The iso-spectral wave functions can be transformed one into another by the Darboux transformation
\begin{equation}\label{psi}
\Psi_{+}=q \left(W+\dfrac{d}{dr_*}\right) \Psi_{-}, \quad q=const,
\end{equation}
so that it is sufficient to calculate quasinormal modes and grey-body factors for only one of the effective potentials. 

\begin{figure}
\resizebox{\linewidth}{!}{\includegraphics*{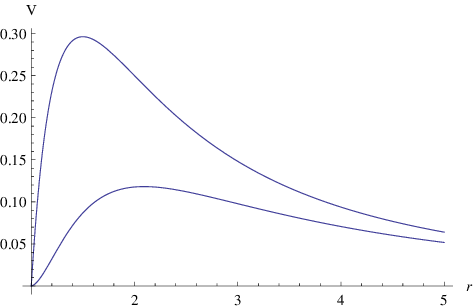}}
\caption{The effective potential for the $\ell=1$ perturbation of the Maxwell field for the Schwarzschild (bottom) and near extreme Bardeen $l_{0}=0.7$ (top) black holes; $r_{0}=1$.}\label{fig:Pot}
\end{figure}

\section{\label{sec:Methods} Methods for finding of quasinormal modes and grey-body factors}

Quasinormal modes represent eigenvalues of the wave-like equations mentioned earlier, corresponding to purely outgoing waves at spatial infinity and purely ingoing waves at the event horizon. Since methods for determining quasinormal frequencies have been extensively discussed in numerous papers, we will provide a brief summary of the three methods employed in this study: the WKB approach with Padé approximants, the Leaver method, and time-domain integration. Furthermore, we will discuss the ordinary WKB method's application in finding grey-body factors.

\subsection{WKB approach}

We will utilize the semi-analytic WKB approach developed by Will and Schutz \cite{Schutz:1985km} to determine the quasinormal modes. The Will-Schutz formula has been extended to higher orders \cite{Iyer:1986np, Konoplya:2003ii, Matyjasek:2017psv} and improved further by implementing Padé approximants \cite{Matyjasek:2017psv, Hatsuda:2019eoj}.

The general form of the WKB formula, as derived by Konoplya \cite{Konoplya:2019hlu}, is as follows:

\begin{eqnarray}
\omega^2&=&V_0+A_2(\K^2)+A_4(\K^2)+A_6(\K^2)+\ldots \\\nonumber
&-& \imo \K\sqrt{-2V_2}\left(1+A_3(\K^2)+A_5(\K^2)+A_7(\K^2)+\ldots\right),
\end{eqnarray}

where $\K=n+1/2$ is a half-integer. The corrections $A_k(\K^2)$, of order $k$, in the eikonal formula are polynomials of $\K^2$ with rational coefficients. These corrections depend on the values of higher derivatives of the potential $V(r)$ at its maximum. To enhance the accuracy of the WKB formula, we will follow the approach of Matyjasek and Opala \cite{Matyjasek:2017psv} by utilizing Padé approximants. Specifically, we will employ the sixth-order WKB method with $\tilde{m}=4,5$, where $\tilde{m}$ is defined in \cite{Matyjasek:2017psv, Konoplya:2019hlu}, as this choice yields the best accuracy in the Schwarzschild limit and is also appropriate for the Bardeen black hole, as confirmed by comparisons with accurate data.

It is important to note that the WKB series converges only asymptotically and does not guarantee improved accuracy at each order. Therefore, it is necessary to cross-validate the WKB results with the convergent Frobenius method. Additionally, the outburst of overtones, occurring when $\ell < n$, cannot be captured by the WKB formula in principle. Hence, in this study, we primarily employ the WKB method as an additional check and solely for calculating the fundamental modes.


\subsection{Frobenius method}

To obtain precise values of quasinormal modes, including overtones with $\ell < n$, we will utilize the Frobenius method, as initially employed by Leaver for calculating quasinormal modes \cite{Leaver:1985ax}. The wave-like equation always exhibits a regular singularity at the horizon $r=r_0$ and an irregular singularity at spatial infinity $r=\infty$. To address this, we introduce a new function:

\begin{equation}\label{reg}
\Psi(r)= P (r, \omega) \left(1-\frac{r_0}{r}\right)^{-\imo\omega/F'(r_0)}y(r),
\end{equation}

Here, the factor $P$ is chosen such that $y(r)$ is regular for $r_0\leq r<\infty$, ensuring that $\Psi(r)$ corresponds to a purely outgoing wave at spatial infinity and a purely ingoing wave at the event horizon. Consequently, we can express $y(r)$ in terms of a Frobenius series:

\begin{equation}\label{Frobenius}
y(r)=\sum_{k=0}^{\infty}a_k\left(1-\frac{r_0}{r}\right)^k.
\end{equation}

By employing Gaussian elimination in the recurrence relation for the expansion coefficients, we reduce the problem to solving an algebraic equation. To enhance convergence speed, we will implement the Nollert improvement \cite{Nollert:1993zz} in its general form for the n-term recurrence relation \cite{Zhidenko:2006rs}.  
When the singular points of the wavelike equation appear within the unit circle $|x| < 1$, we employ a sequence of positive real midpoints as described in  \cite{Rostworowski:2006bp}. 

\subsection{Time domain integration}

To determine the quasinormal modes and asymptotic tails, we will employ the time-domain integration method. We integrate the wave-like equation using the null-cone variables $u=t-r_*$ and $v=t+r_*$. For the discretization scheme, we will apply the Gundlach-Price-Pullin method \cite{Gundlach:1993tp}, which can be expressed as:

$$\Psi\left(N\right)=\Psi\left(W\right)+\Psi\left(E\right)-\Psi\left(S\right) $$
\begin{equation}\label{Discretization}
- \Delta^2V\left(S\right)\frac{\Psi\left(W\right)+\Psi\left(E\right)}{4}+{\cal O}\left(\Delta^4\right).
\end{equation}

Here, the points are denoted as follows: $N\equiv\left(u+\Delta,v+\Delta\right)$, $W\equiv\left(u+\Delta,v\right)$, $E\equiv\left(u,v+\Delta\right)$, and $S\equiv\left(u,v\right)$. Gaussian initial data are imposed on the two null surfaces $u=u_0$ and $v=v_0$. The dominant quasinormal frequencies can be extracted from the time-domain profiles using the Prony method \cite{Konoplya:2011qq}. While extracting the frequency from the time-domain profile with high precision can be challenging, the accuracy for the fundamental mode is guaranteed to be sufficiently good.

\subsection{Scattering problem}

   To investigate Hawking radiation, we will address a different boundary problem that pertains to the scattering of fields in the vicinity of the black hole. Grey-body factors play a crucial role in determining the proportion of initial radiation that gets reflected back to the event horizon by the potential barrier nearby. Subsequently, by employing the Hawking semiclassical formula along with the grey-body factor, we can calculate the amount of radiation that reaches an observer in the far zone. It is worth noting that while temperature is typically the dominant factor in estimating the intensity of Hawking radiation, there are cases where grey-body factors can be equally important \cite{Konoplya:2019ppy}.

We will examine the wave equations under boundary conditions that permit incoming waves from infinity. Due to the scattering symmetry, this is equivalent to considering the scattering of a wave originating from the horizon. The boundary conditions for this scattering problem are as follows:

\begin{equation}\label{BC}
\begin{array}{ccll}
    \Psi &=& e^{-i\omega r_*} + R e^{i\omega r_*},& r_* \to +\infty, \\[5pt]
    \Psi &=& T e^{-i\omega r_*},& r_* \to -\infty, \\
\end{array}
\end{equation}

Here, $R$ and $T$ represent the reflection and transmission coefficients, respectively.

  The effective potentials for test  Maxwell and Dirac fields have the form of  potential barriers which decrease monotonically towards both  infinities \footnote{One of the two effective potentials for the Dirac perturbations have a negative gap, but it is iso-spectral to the other, which is positive definite.}, allowing for application of the WKB approach
  \cite{Schutz:1985km,Iyer:1986np,Konoplya:2003ii}
  to determine $R$ and $T$. As $\omega^2$ is real for the scattering problem, the WKB values for $R$ and $T$ will be
  real as well \cite{Schutz:1985km,Iyer:1986np,Konoplya:2003ii}, and
\begin{equation}\label{1}
		|T|^2 + |R|^2 = 1.
\end{equation}
  Once the reflection coefficient is obtained, we can find the transmission coefficient for each
  multipole number $\ell$,
\begin{equation}
		|{\pazocal A}_{\ell}|^2 = 1 - |R_{\ell}|^2 = |T_{\ell}|^2,
\end{equation}
 where ${\pazocal A}_{\ell}$ is called {\it the grey-body factor}.

  Here we will use the higher order WKB formula \cite{Konoplya:2003ii} for a relatively accurate
  calculation of the grey-body factors. However, this formula is not suitable for very small values of $\omega$,
  which correspond to almost complete wave reflection and have negligible contributions to the overall
  energy emission rate. For this mode, we employed extrapolation of the WKB results at a given order
  to smaller $\omega$ \cite{Inpreparation}. According to \cite{Schutz:1985km,Iyer:1986np}, the reflection coefficient can be
  expressed as follows:
\begin{equation}\label{moderate-omega-wkb}
			R = (1 + e^{- 2 i \pi K})^{-1/2},
\end{equation}
  where $K$ is determined by solving the equation
\begin{equation}
		K - i \frac{(\omega^2 - V_{\max})}{\sqrt{-2 V''_{\max}}}
				- \sum_{i=2}^{i=6} \Lambda_{i}(K) =0,
\end{equation}
  involving the maximum effective potential $V_{\max}$, its second-order derivative $V''_{\max}$
  with respect to the tortoise coordinate, and higher order WKB corrections $\Lambda_i$.
  As mentioned above, the WKB series does not guarantee convergence in each order, but only
  asymptotically, so that usually there is an optimal moderate order at which the accuracy is the best.
  This order depends on the form of the effective
  potential. Here we used 6th order for Maxwell perturbations and 3rd order for the Dirac field with a
  plus-potential, because these orders provide the best accuracy in the Schwarzschild limit, and
  we expect that this will also take place for the Bardeen black hole. This approach was used in a few
  papers \cite{Konoplya:2019hlu,Konoplya:2010vz,Konoplya:2020cbv,Bronnikov:2023uuv}, showing usually reasonably good
  concordance with the numerical integration. Unfortunately, the Pade approximant, which greatly improve the accuracy when finding quasinormal modes,  cannot be used when finding the grey-body factors, because of the non-uniquness of such procedure.

\begin{figure*}
\resizebox{\linewidth}{!}{\includegraphics*{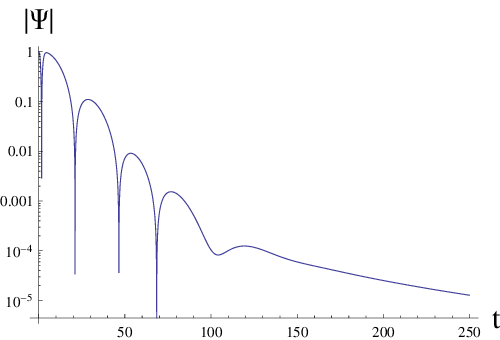}\includegraphics*{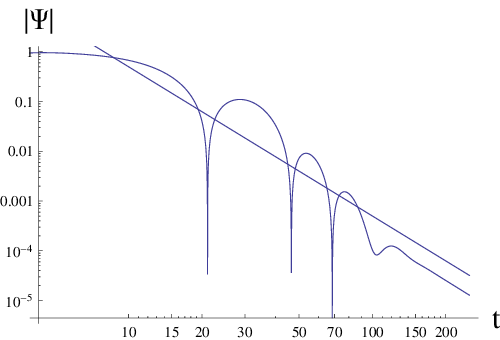}}
\caption{Time-domain profile of the scalar field perturbations $\ell=0$ for the quasi-extremal Bardeen black hole $l_{0} = 0.707107$; $r_{0} =1$. Left: Semi-logarithmic plot; Right: logarithmic plot with the line $\sim t^{-3}$.}
\label{fig:TD}
\end{figure*}
\begin{figure}
	\begin{center}
		\begin{tabular}{c}
			\includegraphics[width=3.2in]{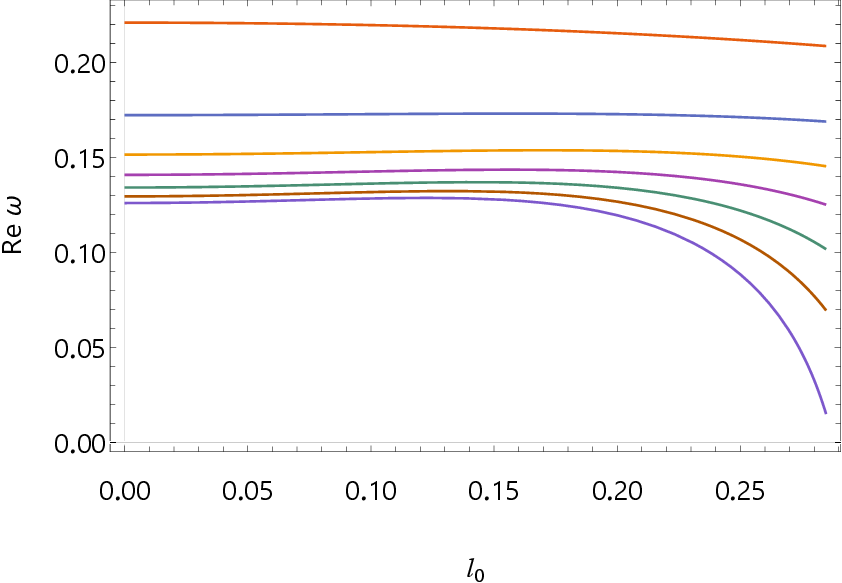}\\
            \includegraphics[width=3.2in]{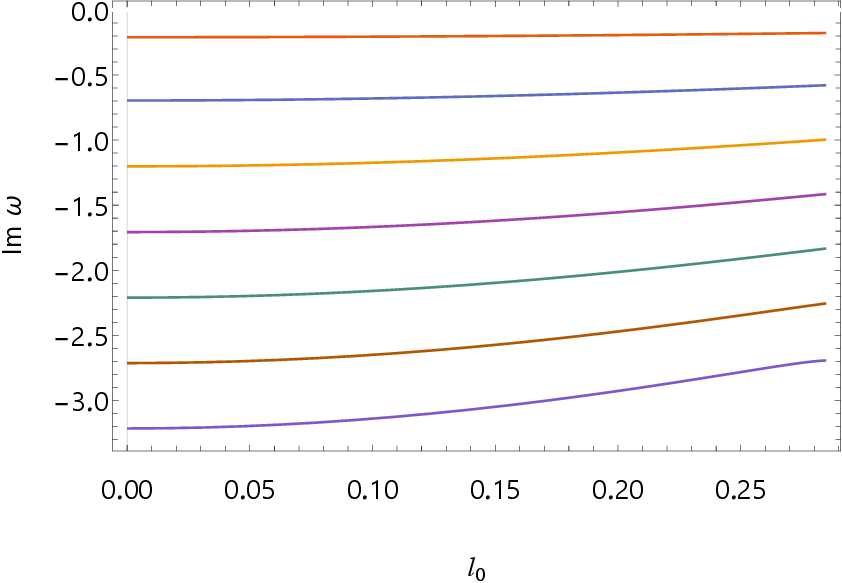}\\
		\end{tabular}
		\caption{Quasinormal frequencies for the scalar perturbations ($\ell=0$) and $n$ from $0$ to $6$.}
		\label{fig:qnm_overtones_0n0}
	\end{center}
\end{figure}

\section{Quasinormal modes and the outburst of overtones}

As a general rule, analytical solutions for quasinormal frequencies of four- and higher-dimensional black holes are not available. However, they can often be obtained in the regime of large multipole numbers $\ell$. By utilizing the first-order WKB formula and expanding the position of the maximum of the effective potential, we can derive the following WKB expression for $\omega$ in terms of $1/L$ and $l_0$, where $L=\ell + \frac{1}{2}$:

\begin{equation}
r_{max} = 3 M-\frac{5
   l_{0}^2}{6 M} -\frac{65 l_{0}^4}{216 M^3} + \mathcal{O}(l_{0}^6)
\end{equation}
$$\omega = \frac{L}{3
   \sqrt{3} M}-\frac{i (2
   n+1)}{6 \sqrt{3} M} 
   +l_{0}^2
   \left(\frac{L}{18 \sqrt{3}
   M^3}+\frac{i (2 n+1)}{54
   \sqrt{3}
   M^3}\right)+$$
\begin{equation}
l_{0}^4 \left(\frac{17 L}{648
   \sqrt{3} M^5}+\frac{7 i (2
   n+1)}{324 \sqrt{3}
   M^5}\right)+ \mathcal{O} \left(\frac{1}{L}, l_{0}^6\right).
\end{equation}

It is evident that the above expressions are accurate in the limit of large $\ell$, while still providing reasonable accuracy even at moderate values of $\ell$. When $l_0 = 0$, these analytical formulas reduce to the well-known expressions for the Schwarzschild black hole \cite{Ferrari:1984zz} (see \cite{Zhidenko:2003wq,Churilova:2021nnc} and references therein for various generalizations of these relations). Furthermore, the aforementioned eikonal formula can be derived through the correspondence between eikonal quasinormal modes and null geodesics \cite{Cardoso:2008bp,Konoplya:2017wot,Konoplya:2022gjp}. However, it should be noted that this correspondence has several limitations, as described in \cite{Konoplya:2017wot,Konoplya:2022gjp,Konoplya:2019hml}.

From this point onwards in this section, we will compute quasinormal frequencies in units of the fixed radius of the event horizon, $r_0=1$, rather than fixing the mass of the black hole, $M$. The mass and radius of the event horizon are related by the following equation:

\begin{equation}
M = \frac{(r_0^2 + l_0^2)^{3/2}}{2 r_0^2}.
\end{equation}

This equation allows us to switch between the units $r_0=1$ and $M=1$. The units of the fixed event horizon are more convenient for the application of the Frobenius method, which requires a precise understanding of the singular points of the differential equation under consideration.

\begin{figure}
	\begin{center}
		\begin{tabular}{c}
			\includegraphics[width=3.2in]{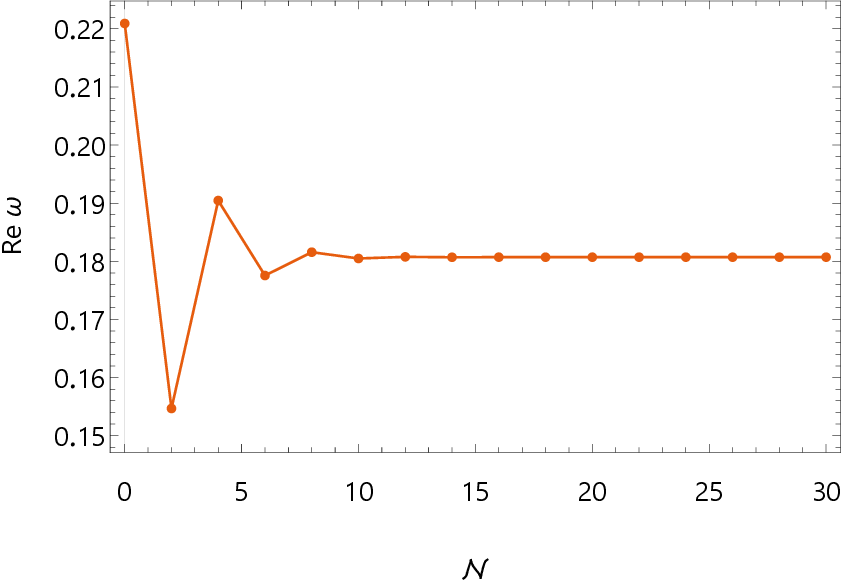}\\
            \includegraphics[width=3.2in]{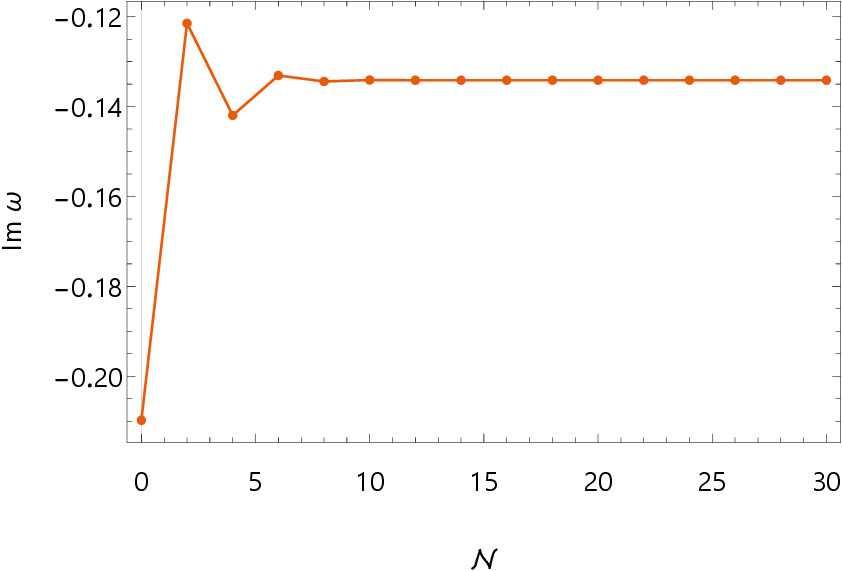}\\
		\end{tabular}
		\caption{An example of the convergence of the scalar field QNMs $\omega$ calculated by the Frobenius method as the order of the expansion $\mathcal{N}$ in terms of small $l_{0}$ is increased ($n=0$, $\ell=0$) and $l_{0}=0.46669$.}
		\label{fig:converg_frob_poly}
	\end{center}
\end{figure}

The fundamental quasinormal modes for the test scalar field ($\ell=n=0$), electromagnetic field ($\ell=1$, $n=0$), and Dirac field ($\ell=1/2$, $n=0$) are presented in Tables I-III. The accurate values are obtained using the convergent Frobenius method \cite{Leaver:1985ax} for scalar and electromagnetic fields, while for the Dirac field, which has a non-polynomial wave equation, we were limited to using the WKB method. A comparison of these values with earlier published results \cite{Fernando:2012yw,Toshmatov:2015wga} reveals that the ordinary 6th order WKB data (without Padé approximants) is inaccurate, with a relative error reaching approximately $10\%$. However, the advanced WKB method with Padé approximants considered here \cite{Matyjasek:2017psv} shows much better agreement with accurate quasinormal frequencies, reducing the relative error to a small fraction of one percent. The Frobenius method can only be applied to polynomial forms of the metric function and effective potentials. To meet this requirement, we expand the metric function in terms of small $l_0$ (see Appendix for the explicit form of the expansion) and apply the Frobenius method to this expanded metric. We observe that the quasinormal frequencies calculated using the metric function expanded up to the 20th and 24th orders in $l_0$ are in excellent agreement (see Table I-II). An example of the convergence in terms of $l_0$ is shown in Figure \ref{fig:converg_frob_poly}. The higher the overtone $n$ one needs to find, the higher the order of expansion of the metric that should be used.

Integration of the wave equation in the time domain and subsequent extraction of dominant frequencies from the profile using the Prony method also confirm the accuracy of the Frobenius and WKB (with Padé) data. For instance, for the quasi-extremal state $\ell_0 = 0.707107$ and $\ell=0$ scalar perturbations, the frequency $\omega = 0.12958 - 0.09393i$ is extracted from the time-domain profile. This is in close agreement with the Frobenius method for the metric expanded up to the 24th order, which gives $\omega = 0.129521 - 0.0939448i$, and the 6th order WKB method with Padé approximants, which gives $\omega = 0.129879 - 0.093637i$. On the other hand, the usual 6th order WKB method produces an approximately $10\%$ error in the damping rate.

At asymptotically late times, when the quasinormal ringing is suppressed by power-law tails, the Price decay law for the test scalar and gravitational fields \cite{Price:1972pw} is fulfilled:

\begin{equation}
|\Psi| \sim t^{-(2\ell +3)}, \quad t \rightarrow \infty.
\end{equation}

This can be observed in Figure \ref{fig:TD}.

The first several overtones exhibit the most interesting feature. They deviate from their Schwarzschild values at a significantly higher rate compared to the fundamental mode, and this rate increases with the overtone number (see Figure \ref{fig:qnm_overtones_0n0}). While the fundamental mode undergoes only a slight change, limited to moderate values of the quantum deformation parameter $l_{0}$ (as shown in Figure \ref{fig:qnm_overtones_0n0}), the real oscillation frequency of the sixth overtone decreases by more than twice its Schwarzschild limit.

This outburst of overtones occurs because the Bardeen spacetime differs from the Schwarzschild spacetime mainly in a region near the event horizon. This characteristic could potentially enable us to discern the quantum corrections near the horizon through the early phase of the quasinormal ringing, which includes contributions from the overtones \cite{Konoplya:2022pbc,Giesler:2019uxc}. It is important to note that if we consider the Bardeen spacetime as a deformation of the Schwarzschild black hole, this deformation is not highly localized and, therefore, is not related to the so-called "high-frequency perturbations" (which actually refer to "highly localized deformation") discussed in \cite{Nollert:1996rf} and subsequent works. The deformation of the Schwarzschild spacetime introduced by the parameter $l_{0}$ is smooth and extends over a neighborhood of the event horizon, gradually decaying as we move away from the black hole.

\begin{table*}
\caption{\label{tab:qnm_freq_000} Fundamental quasinormal modes for the scalar perturbations ($n=0$, $\ell=0$).}
\begin{ruledtabular}
\begin{tabular}{c|ccc|cc}
        &\multicolumn{3}{c|}{WKB}                                                                        &\multicolumn{2}{c}{Frobenius}\\
$l_{0}$ &       6th                     &       6th (Pade 5)            &       6th (Pade 6)            &       20th order              &       24th order              \\ \hline
0	    &	$	0.220934-0.201633 i	$	&	$	0.221584-0.209367 i	$	&	$	0.221357-0.208847 i	$	&	$	0.220910-0.209791 i	$	&	$	0.220910-0.209791 i	$	\\
0.07698	&	$	0.220288-0.199177 i	$	&	$	0.220818-0.206828 i	$	&	$	0.220585-0.206324 i	$	&	$	0.220146-0.207248 i	$	&	$	0.220146-0.207248 i	$	\\
0.15396	&	$	0.218078-0.192489 i	$	&	$	0.218398-0.199312 i	$	&	$	0.218146-0.198876 i	$	&	$	0.217717-0.199765 i	$	&	$	0.217717-0.199765 i	$	\\
0.23094	&	$	0.214096-0.182285 i	$	&	$	0.213780-0.187430 i	$	&	$	0.213541-0.187065 i	$	&	$	0.213239-0.187801 i	$	&	$	0.213239-0.187801 i	$	\\
0.30792	&	$	0.209605-0.165605 i	$	&	$	0.205935-0.172526 i	$	&	$	0.205928-0.172054 i	$	&	$	0.206151-0.172156 i	$	&	$	0.206151-0.172156 i	$	\\
0.3849	&	$	0.203168-0.140512 i	$	&	$	0.195809-0.154749 i	$	&	$	0.195900-0.154227 i	$	&	$	0.195821-0.154042 i	$	&	$	0.195821-0.154042 i	$	\\
0.46188	&	$	0.186592-0.119218 i	$	&	$	0.183120-0.134872 i	$	&	$	0.182642-0.134402 i	$	&	$	0.181717-0.135270 i	$	&	$	0.181717-0.135270 i	$	\\
0.53886	&	$	0.166413-0.106640 i	$	&	$	0.164673-0.118464 i	$	&	$	0.164432-0.118343 i	$	&	$	0.164121-0.118732 i	$	&	$	0.164119-0.118735 i	$	\\
0.61584	&	$	0.149688-0.096004 i	$	&	$	0.147642-0.106578 i	$	&	$	0.147526-0.106419 i	$	&	$	0.146968-0.106725 i	$	&	$	0.146995-0.106719 i	$	\\
0.69282	&	$	0.134525-0.086147 i	$	&	$	0.132643-0.095711 i	$	&	$	0.132524-0.095546 i	$	&	$	0.132241-0.0959643 i$	&	$	0.132129-0.0958409 i$	\\
0.707107&	$	0.131840-0.084424 i	$	&	$	0.129995-0.093798 i	$	&	$	0.129879-0.093637 i	$	&	$	0.129702-0.0941100 i$	&	$	0.129521-0.0939448 i$	\\
\end{tabular}
\end{ruledtabular}
\end{table*}

\begin{table*}
\caption{\label{tab:qnm_freq_101} Fundamtenal quasinormal modes for the electromagnetic perturbations  ($n=0$, $\ell=1$).}
\begin{ruledtabular}
\begin{tabular}{c|ccc|cc}
        &\multicolumn{3}{c|}{WKB}                                                                        &\multicolumn{2}{c}{Frobenius}\\
$l_{0}$     &       6th                     &       6th (Pade 5)            &       6th (Pade 6)            &       20th order              &       24th order              \\ \hline
0         	&	$	0.496383-0.185274 i	$	&	$	0.496509-0.184993 i	$	&	$	0.496509-0.184961 i	$	&	$	0.496527-0.184975 i	$	&	$	0.496527-0.184975 i	$	\\
0.069282	&	$	0.494945-0.183682 i	$	&	$	0.495092-0.183337 i	$	&	$	0.495086-0.183305 i	$	&	$	0.495104-0.183318 i	$	&	$	0.495104-0.183318 i	$	\\
0.138564	&	$	0.490564-0.178948 i	$	&	$	0.490767-0.178429 i	$	&	$	0.490754-0.178398 i	$	&	$	0.490772-0.178411 i	$	&	$	0.490772-0.178411 i	$	\\
0.207846	&	$	0.483055-0.171211 i	$	&	$	0.483353-0.170467 i	$	&	$	0.483338-0.170434 i	$	&	$	0.483353-0.170452 i	$	&	$	0.483353-0.170452 i	$	\\
0.277128	&	$	0.472162-0.160742 i	$	&	$	0.472577-0.159796 i	$	&	$	0.472563-0.159759 i	$	&	$	0.472572-0.159784 i	$	&	$	0.472572-0.159784 i	$	\\
0.34641	    &	$	0.457592-0.147991 i	$	&	$	0.458107-0.146925 i	$	&	$	0.458095-0.146889 i	$	&	$	0.458099-0.146918 i	$	&	$	0.458099-0.146918 i	$	\\
0.415692	&	$	0.439053-0.133649 i	$	&	$	0.439614-0.132573 i	$	&	$	0.439603-0.132545 i	$	&	$	0.439605-0.132570 i	$	&	$	0.439605-0.132570 i	$	\\
0.484974	&	$	0.416343-0.118720 i	$	&	$	0.416890-0.117721 i	$	&	$	0.416882-0.117704 i	$	&	$	0.416881-0.117720 i	$	&	$	0.416881-0.117720 i	$	\\
0.554256	&	$	0.389608-0.104530 i	$	&	$	0.390103-0.103633 i	$	&	$	0.390099-0.103626 i	$	&	$	0.390100-0.103635 i	$	&	$	0.390096-0.103636 i	$	\\
0.623538	&	$	0.359764-0.092365 i	$	&	$	0.360192-0.091561 i	$	&	$	0.360190-0.091559 i	$	&	$	0.360256-0.0915651 i$	&	$	0.360201-0.0915712 i$	\\
0.69282 	&	$	0.328400-0.082759 i	$	&	$	0.328778-0.082042 i	$	&	$	0.328774-0.082038 i	$	&	$	0.329444-0.0819892 i$	&	$	0.328945-0.0820359 i$	\\
0.707107	&	$	0.321891-0.081076 i	$	&	$	0.322262-0.080374 i	$	&	$	0.322258-0.080369 i	$	&	$	0.323296-0.0802851 i$	&	$	0.322542-0.0803561 i$	\\
\end{tabular}
\end{ruledtabular}
\end{table*}

\begin{figure}
	\begin{center}
		\begin{tabular}{c}
			\includegraphics[width=3.2in]{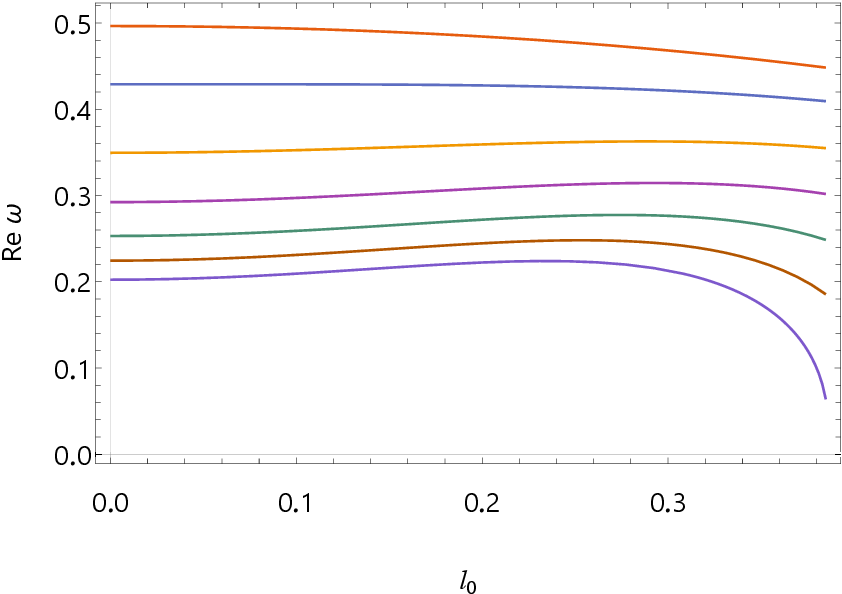}\\
            \includegraphics[width=3.2in]{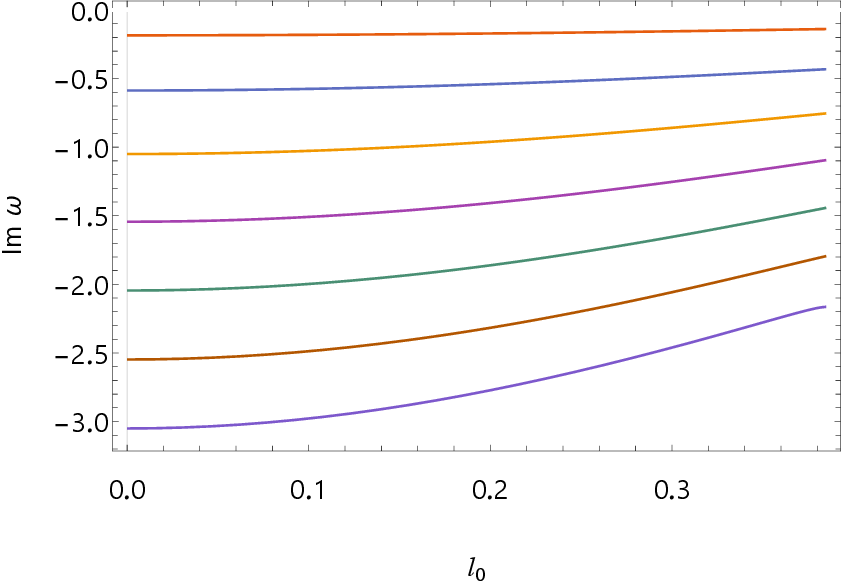}\\
		\end{tabular}
		\caption{Quasinormal frequencies for the electromagnetic perturbations  ($\ell=1$) and $n$ from $0$ to $6$.}
		\label{fig:qnm_overtones_1n1}
	\end{center}
\end{figure}

\begin{table*}
\caption{\label{tab:qnm_freq_12012} Fundamental quasinormal modes for the Dirac perturbations ($n=0$, $\ell=1/2$).}
\begin{ruledtabular}
\begin{tabular}{c|ccc}
        &\multicolumn{3}{c}{WKB}\\                                                                       
$l_{0}$     &       6th                     &       6th (Pade 5)            &       6th (Pade 6)            \\ \hline
0	        &	$	0.366162-0.194105 i	$	&	$	0.365943-0.193956 i	$	&	$	0.365813-0.193991 i	$	\\
0.07698	    &	$	0.364600-0.191742 i	$	&	$	0.364392-0.191630 i	$	&	$	0.364267-0.191664 i	$	\\
0.15396	    &	$	0.359889-0.184810 i	$	&	$	0.359702-0.184821 i	$	&	$	0.359589-0.184851 i	$	\\
0.23094	    &	$	0.351950-0.173766 i	$	&	$	0.351761-0.174019 i	$	&	$	0.351660-0.174036 i	$	\\
0.30792	    &	$	0.340642-0.159328 i	$	&	$	0.340367-0.159983 i	$	&	$	0.340279-0.159978 i	$	\\
0.3849	    &	$	0.325739-0.142490 i	$	&	$	0.325228-0.143677 i	$	&	$	0.325166-0.143648 i	$	\\
0.46188	    &	$	0.306947-0.124700 i	$	&	$	0.306169-0.126350 i	$	&	$	0.306135-0.126322 i	$	\\
0.53886	    &	$	0.284334-0.107994 i	$	&	$	0.283432-0.109828 i	$	&	$	0.283418-0.109814 i	$	\\
0.61584	    &	$	0.259273-0.094155 i	$	&	$	0.258320-0.095939 i	$	&	$	0.258314-0.095930 i	$	\\
0.69282	    &	$	0.233822-0.083456 i	$	&	$	0.232929-0.085144 i	$	&	$	0.232909-0.085107 i	$	\\
0.707107    &	$	0.229173-0.081764 i	$	&	$	0.228298-0.083422 i	$	&	$	0.228278-0.083383 i	$	\\
\end{tabular}
\end{ruledtabular}
\end{table*}

The significant decrease in the real oscillation frequencies of the overtones with increasing deformation parameter $l_{0}$ suggests the possibility of purely imaginary quasinormal modes in the spectrum at larger values of $l_{0}$. However, due to the computational limitations, we were unable to find such modes as they would require much longer computation times. It is important to note that the quantum correction parameter is intended to represent a relatively small or at most moderate correction to the Schwarzschild spacetime. Therefore, the metric should remain sufficiently far from its extreme state, ensuring that the corrections remain within a reasonable range.

\begin{figure*}
\resizebox{\linewidth}{!}{\includegraphics*{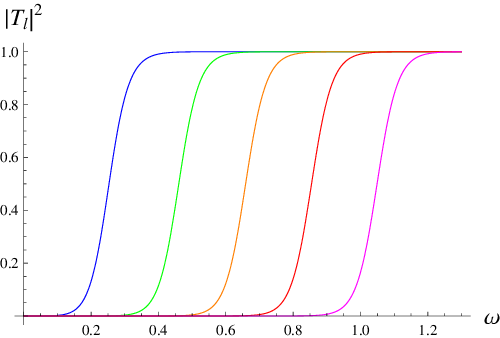}\includegraphics*{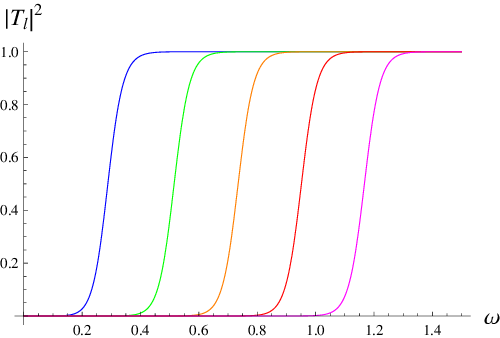}}
\caption{Grey-body factors for the Maxwell field as function of $\omega$ for the Schwarzschild solution (left panel) and the near extreme ($l_{0} =0.7$) Bardeen black hole (right panel) computed by the 6th order WKB approach for the first five multipoles $\ell =1, 2,..5$ from left to right.}\label{fig:T}
\end{figure*}

\begin{figure*}
\resizebox{\linewidth}{!}{\includegraphics*{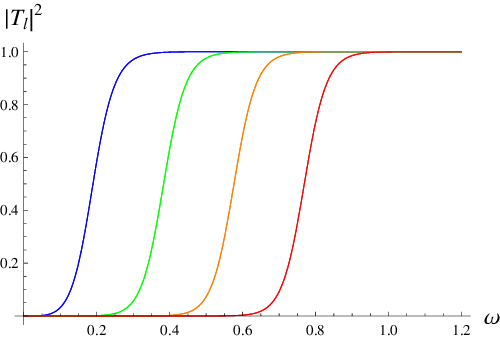}\includegraphics*{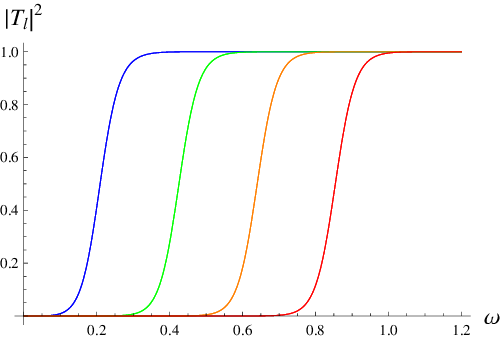}}
\caption{Grey-body factors for the Dirac field as function of $\omega$ for the Schwarzschild solution (left panel) and the near extreme ($l_{0} =0.7$) Bardeen black hole (right panel) computed by the 3d order WKB approach for the first five multipoles $\ell =1, 2,..5$ from left to right.}
\end{figure*}

\begin{figure*}
\resizebox{\linewidth}{!}{\includegraphics*{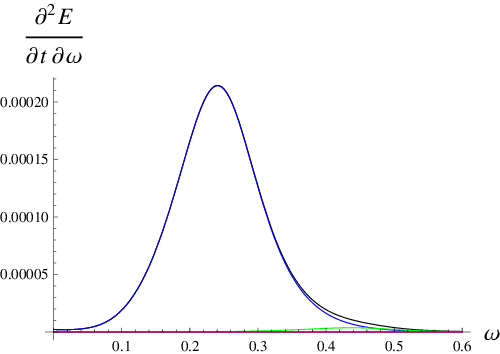}\includegraphics*{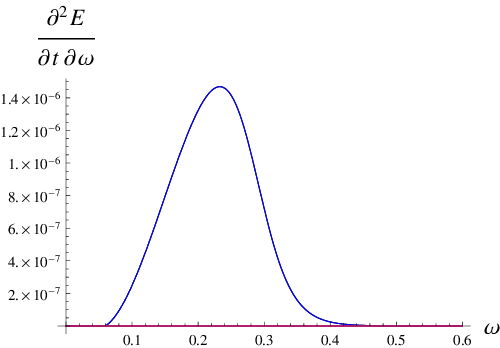}}
\caption{Energy emission rates per unit of frequency for the Maxwell field for the Schwarzschild spacetime (left plot) and for the near extreme Bardeen black hole $l_{0}=0.7$ (right plot). One can see that the contribution consists almost completely from the first multipole $\ell=1$ (blue, top) and $\ell=2$ (yellow, bottom), while higher multipoles could neglected. The black line is for the total emission rate summed over the first five multipoles.}\label{fig:PerTime}
\end{figure*}

\section{Hawking radiation}

  It is essential to note that radiation from test fields surpasses that of gravitons: as is known using
  Schwarzschild black holes as an example, gravitons account for less
  than $2\%$ of the total radiation flux, as illustrated in \cite{Page:1976df} and also summarized in
  Table I of \cite{Konoplya:2019ppy}.

  Here we will assume that the black hole is in a state of thermal
  equilibrium with its surroundings. This means that the temperature of the black hole remains constant
  between the emission of two consecutive particles. According to this assumption, the system can be
  characterized by the canonical ensemble, which is extensively discussed in the literature (see,
  e.g., \cite{Kanti:2004nr}). Consequently, the well-known Hawking formula for the energy emission rate 
  \cite{Hawking:1975vcx} can be applied:
\begin{align}\label{energy-emission-rate}
		\frac{\text{d}E}{\text{d} t} = \sum_{\ell}^{} N_{\ell}
		| \pazocal{A}_\ell |^2 \frac{\omega}		
				{\exp\left(\omega/T_\text{H}\right)\pm1}\cdot \frac{\text{d} \omega}{2 \pi},
\end{align}
  where $T_H$ is the Hawking temperature, $\pazocal{A}_\ell$ are grey-body factors, and $N_\ell$
  are the multiplicity factors which depend on the number of species of particles and $\ell$.
  The Hawking temperature is \cite{Hawking:1975vcx}
\begin{equation}
T = \left.\frac{f'(r)}{(4 \pi)}\right\vert_{r=r_{0}}. 
\end{equation}
The dependence of the temperature on the parameter $l_{0}$ is shown in fig. \ref{fig:T}.

\begin{figure}
\resizebox{\linewidth}{!}{\includegraphics*{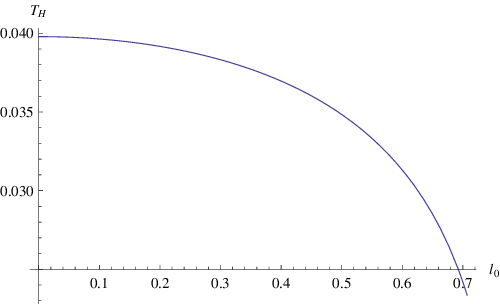}}
\caption{Temperature of the Bardeen black hole at fixed mass $M=1$ as a function of the deforamtion parameter $l_{0}$.}\label{fig:Tem}
\end{figure}

  As long as we are considering the evaporation stage when the black hole is large enough, one can
  neglect emission of massive particles \cite{Gaina:1985xg}  as compared to massless ones. Accordingly, we will here consider
  only massless particles of the Standard Model, and, as was argued above, we neglect the
  emission of gravitons.
  Once the black hole is small enough, or, alternatively, the deformation parameter $l_{0}/r_{0}$ is large enough,
  massive particles, such as electrons and positrons, are emitted in an ultrarelativistic mode, that is,
  roughly at the same rate as massless neutrinos for each helicity.

  The multiplicity factors for the 4D spherically symmetrical black holes consist of the number of degenerate
  $m$-modes (which are $m = -\ell, -\ell+1, ....-1, 0, 1, ...\ell$, that is  $2 \ell +1$ modes) multiplied by the
  number of species of particles, which in turn depends also on the number of polarizations and helicities
  of particles. Therefore, we have
\begin{eqnarray}
		N_{\ell} = 2 (2 \ell+1) \qquad (\rm Maxwell),
\\[5pt] 
		N_{\ell} = 8 \ell \qquad (\rm Dirac),
\end{eqnarray}
  The multiplicity factor for the Dirac field is fixed taking into account both the ``plus'' and ``minus''
  potentials related by Darboux transformations, which leads to an isospectral problem and the same
  grey-body factors for both chiralities. We will use here the ``plus'' potential, because the WKB results
  are more accurate for that case in the Schwarzschild limit.

  The energy emitted makes the black hole mass decrease at the following rate \cite{Page:1976df}:
\begin{equation}        \label{M-dot}
		\frac{d M}{d t} = -\frac{\hbar c^4}{G^2} \frac{\alpha_{0}}{M^2}
		\approx - 1.91\times 10^{29}\alpha_0\ \frac{\rm g}{\rm s}
			\cdot\bigg(\frac{\rm g^2}{M_0^2}\bigg),
\end{equation}
  where we have restored the dimensional constants. Here $\alpha_{0} = d E/d t$ is taken
  in units used in Table IV for a given initial mass $M_0$.  Indeed, this can be seen by comparison of
  our data in the Schwarzschild limit with \cite{Page:1976df}. A relatively small difference with the Page's
  results (less than two percents) is due to the systematic error of the WKB method \cite{Konoplya:2019hlu} and can be neglected, taking into account that the overall effect of deviation of the energy emission rate from the Schwarzschild limit achieves a few orders, i.e. the relative error is a few orders smaller than the effect.

\begin{table}
\begin{tabular}{|c|c|c|}
  \hline
   \hline
  $l_{0}/M$ & Maxwell & Dirac   \\
\hline
  0 ($\ell=1$) &  $0.0000335107$ &  $0.00015963$  \\
  0 ($\ell=2$) &  $6.67916 \cdot 10^{-7}$ &  $5.8691 \cdot 10^{-6}$  \\
  0 ($\ell=3$) &  $1.00693 \cdot 10^{-8}$ &  $1.16695 \cdot 10^{-7}$ \\
\hline
  0 (total) & $0.0000341888$  &  $0.000165617$ \\
\hline
  0.1 ($\ell=1$) &  $0.0000323629$ &  $0.000155663$ \\
  0.1 ($\ell=2$) &  $6.28544 \cdot 10^{-7}$ &  $5.58669 \cdot 10^{-6}$ \\
  0.1 ($\ell=3$) &  $9.23148 \cdot 10^{-9}$ &  $1.08274 \cdot 10^{-7}$\\
\hline
  0.1 (total) &   $0.0000330008$  &  $0.00016136$  \\
\hline
  0.2 ($\ell=1$) &  $0.0000290019$ &  $0.000144034$ \\
  0.2 ($\ell=2$) &  $5.1916 \cdot 10^{-7}$ &  $4.78308  \cdot 10^{-6}$  \\
  0.2 ($\ell=3$) &  $7.02324 \cdot 10^{-9}$ &  $8.55366 \cdot 10^{-8}$  \\
\hline
  0.2 (total) &  $0.0000295282$ &  $0.000148904$  \\
  \hline
  0.3 ($\ell=1$) &  $0.0000236953$ &  $0.00012464$  \\
  0.3 ($\ell=2$) & $3.64925 \cdot 10^{-7}$ &  $3.5915 \cdot 10^{-6}$  \\
  0.3 ($\ell=3$) &  $4.24242 \cdot 10^{-9}$ &  $5.53788 \cdot 10^{-8}$  \\
\hline
  0.3 (total) & $0.0000240645$ &  $0.000128288$  \\
  \hline
  0.4 ($\ell=1$) & $0.0000169636$ & $0.000102356$  \\
  0.4 ($\ell=2$) & $2.03653 \cdot 10^{-7} $ & $2.23439 \cdot 10^{-6}$  \\
  0.4 ($\ell=3$) & $1.84218 \cdot 10^{-9}$ &  $2.70179 \cdot 10^{-8}$  \\
\hline
  0.4 (total) &   $0.0000171691$ &  $0.000104618$ \\
  \hline
  0.5 ($\ell=1$) &  $9.71567 \cdot 10^{-6}$ &  $0.0000658481$  \\
  0.5 ($\ell=2$) &  $7.69549 \cdot 10^{-8} $ &  $1.01073 \cdot 10^{-6}  $  \\
  0.5 ($\ell=3$) &  $4.57759 \cdot 10^{-10}$ &  $8.10308 \cdot 10^{-9}$  \\
\hline
  0.5 (total) &  $9.79308 \cdot 10^{-6}$ &  $0.000066867$  \\
  \hline
  0.6 ($\ell=1$) &  $3.45285 \cdot 10^{-6}$ &  $0.000031204$  \\
  0.6 ($\ell=2$) &  $1.26319 \cdot 10^{-8}$ &  $2.34735 \cdot 10^{-7}$  \\
  0.6 ($\ell=3$) &  $3.44314 \cdot 10^{-11}$ &  $8.70239 \cdot 10^{-10}$  \\
\hline
  0.6 (total) &  $3.46552  \cdot 10^{-6}$ &  $0.00003144$  \\
  \hline
  0.7 ($\ell=1$) &  $2.34399 \cdot 10^{-7}$ &  $4.29524 \cdot 10^{-6}$  \\
  0.7 ($\ell=2$) &  $1.18349 \cdot 10^{-10}$ &  $4.79822 \cdot 10^{-9}$  \\
  0.7 ($\ell=3$) &  $4.36355 \cdot 10^{-14}$ &  $2.55798  \cdot 10^{-12}$  \\
\hline
  0.7 (total) &  $2.34517\cdot 10^{-7}$ &  $4.30004 \cdot 10^{-6}$  \\
  \hline
   \hline
\end{tabular}
\caption{Energy emission rate $d E/ d t$ for the Bardeen black hole}
\label{TableIV}
\end{table}


From Table \ref{TableIV}, it is evident that the energy emission rate $dE/dt$ decreases by three orders of magnitude as the deformation parameter $l_{0}$ approaches the quasi-extreme limit. An opposite effect has been observed in another model of quantum correction \cite{Bronnikov:2023uuv}, based on the non-perturbative Kazakov-Solodukhin solution \cite{Kazakov:1993ha}, where the Hawking temperature of the quantum corrected black hole is the same as for the Schwarzschild case. The effect observed in our current research can be attributed to two factors. The primary and dominant factor is the decrease in the Hawking temperature of the black hole when the quantum correction is introduced (see fig. \ref{fig:T}). The second factor is the grey-body coefficient. Analyzing the shape of the effective potentials of the Bardeen black hole shown in fig. \ref{fig:Pot}, it is evident that the height of the potential barrier increases with the inclusion of the quantum correction. Consequently, a taller effective potential reflects a larger portion of the initial radiation flow, contributing to the suppression of Hawking radiation.

From fig. \ref{fig:PerTime}, we observe that while the emission rate of the Maxwell field per unit frequency decreases significantly when the near-extreme quantum correction is applied, the peak of the radiation occurs at approximately the same frequency $\omega$. Similar behavior is observed for the Dirac particles.

We have calculated the contributions of the first five multipoles $\ell$. However, it is noteworthy that the contributions of only the first two multipoles are significant for the total energy emission rate. Therefore, higher multipoles can be safely neglected.

\section{Conclusions}

The Bardeen spacetime characterizes a regular black hole featuring a de Sitter core in place of a central singularity. By interpreting this spacetime as the metric of a quantum-corrected black hole within the framework of string T-duality \cite{Nicolini:2019irw}, we conducted a comprehensive analysis of the quasinormal spectrum for scalar, electromagnetic, and Dirac fields in this background. Our investigation revealed that some previous studies focusing on the fundamental mode exhibited inaccuracies, allowing for a relative error of up to $10\%$. However, our primary findings pertain to the calculations of overtones, which probe the geometry near the event horizon and thus play a crucial role in discerning potential quantum corrections through observations of classical radiation emitted by black holes. Notably, we observed that the initial few overtones deviate from their Schwarzschild values at a significantly higher rate compared to the fundamental mode. This deviation stems from the quantum correction-induced deformation of the Schwarzschild spacetime.

Additionally, we derived an analytical expression for the quasinormal modes in the eikonal regime and established the decay law at asymptotically late times.

Furthermore, we computed the grey-body factors and energy emission rates for massless fields surrounding the Bardeen black hole. Our analysis revealed that the quantum correction leads to a significant suppression of the Hawking radiation intensity by several orders of magnitude. This suppression arises from two factors: a decrease in the Hawking temperature and a smaller grey-body factor.

Our study on the Hawking evaporation of the Bardeen black hole holds potential significance in estimating the number of primordial black holes formed during the early Universe. Specifically, for $M \lesssim 10^{15} g$, the evaporation of primordial black holes is predominantly constrained by big bang nucleosynthesis \cite{Carr:2020gox}. Therefore, the quantum correction examined in this study could serve as one of the mechanisms, alongside those considered in \cite{Franciolini:2023osw}, that chnages the speed of the black hole evaporation, potentially influencing the constraints imposed by big bang nucleosynthesis on the abundance of primordial black holes.

There are several avenues for extending our work. Firstly, one could investigate the Hawking radiation of massive particles to obtain a more accurate understanding of the radiation process during the later stages of black hole evaporation. Secondly, the decreasing real oscillation frequencies of the overtones at moderate values of $l_{0}$ suggest that, at a sufficiently large deformation parameter, some of these overtones may become purely imaginary, corresponding to non-oscillatory modes. However, due to the slow convergence of the procedure in this regime, we were unable to explore this phenomenon within a reasonable computation time. It is possible that alternative methods, such as the Bernstein polynomial method \cite{Fortuna:2020obg, Konoplya:2022zav}, which are more effective at finding purely imaginary modes, could offer a solution to this challenge.

\section{Appendix}
The expansion of the metric function $f(r)$ in terms of small $l_{0}$ has the following form:

\begin{eqnarray}
        f(r)& =&\left(1-\frac{2 M}{r}\right)+\frac{3 l_0^2 M}{r^3}-\frac{15 l_0^4 M}{4r^5}+\frac{35 l_0^6 M}{8 r^7}\nonumber \\
        &-&\frac{315 l_0^8 M}{64 r^9}+\frac{693 l_0^{10}M}{128 r^{11}}-\frac{3003 l_0^{12} M}{512 r^{13}}+\frac{6435 l_0^{14} M}{1024r^{15}}\nonumber \\
        &-&\frac{109395 l_0^{16} M}{16384 r^{17}}+\frac{230945 l_0^{18} M}{32768r^{19}}-\frac{969969 l_0^{20} M}{131072 r^{21}}\nonumber \\
        &+&\frac{2028117 l_0^{22} M}{262144r^{23}}-\frac{16900975 l_0^{24} M}{2097152 r^{25}}+O\left(l_0^{26}\right).
\end{eqnarray}

\acknowledgments{The authors would like to acknowledge useful discussions with Alexander Zhidenko. They would also like to acknowledge the support of the Research Centre for Theoretical Physics and Astrophysics, Institute of Physics, Silesian University in Opava. D. O. thanks the financial support of the internal grant of the Silesian University in Opava SGS/30/2023. B.~A. acknowledges support by grants F-FA-
2021-432, and MRB-2021-527 from the Agency for Innovative Development of Uzbekistan and Erasmus+ project towards his stay at the Silesian University in Opava.}

\bibliography{bardeen_bh_qnms}

\end{document}